# Observation of resistive switching and diode effect in the conductivity of TiTe$_2$ point contacts

O. E. Kvitnitskaya[1,2], L. Harnagea[2,3], O.D. Feia[4,5,2], D. V. Efremov[2], B. Büchner[2,6], Yu. G. Naidyuk[1]

[1]*B. Verkin Institute for Low Temperature Physics and Engineering, NAS of Ukraine, 61103 Kharkiv, Ukraine*

[2]*Leibniz Institute for Solid State and Materials Research, IFW Dresden, 01069 Dresden, Germany*

[3]*I-HUB Quantum Technology Foundation, Indian Institute of Science Education and Research (IISER), Pune 411008, India*

[4]*Kyiv Academic University, 03142, Kyiv, Ukraine*

[5]*G.V. Kurdyumov Institute for Metal Physics, NAS of Ukraine, 03142, Kyiv, Ukraine*

[6]*Institute of Solid State and Materials Physics and Würzburg-Dresden Cluster of Excellence ct.qmat, Technische Universität Dresden, 01062 Dresden, Germany*

**Abstract**

We measured the *I(V)* and *dV/dI(V)* characteristics of TiTe$_2$-based point contacts (PCs) from room to helium temperatures. Features indicating the emergence of a charge density wave (CDW) were detected. They represent symmetrical relatively V=0 maxima in *dV/dI(V)* around +/- 150 mV at liquid helium temperatures, which disappear above 150 K, similar to the case of sister CDW compound TiSe$_2$. Applying higher voltages above 200 mV, we observed resistive switching in TiTe$_2$ PCs from a metallic-like low-resistance state to non-metallic type high-resistance state with a change of resistance by an order of magnitude. A unique diode-like effect was registered in "soft" TiTe$_2$ PCs with hysteretic *I(V)* at negative voltage on TiTe$_2$. Discovering the resistive switching and diode effect adds TiTe$_2$ to the transition-metal dichalcogenides, which could be useful in developing non-volatile ReRAM and other upcoming nanotechnologies.

**Introduction**

Layered transition-metal dichalcogenides (TMDs) with weak van der Waals bonding exhibit a variety of physical features that hold promise for the development of novel nanoelectronic devices and components. Among this class of materials, titanium dichalcogenides $TiX_2$ (X = S, Se, Te) display a variety of physical phenomena, depending on their lattice distortions, native defects, self-doping, intercalation, and external pressure. They can display semiconducting, semimetallic, charge density wave (CDW), and superconducting behavior, which together with their layered structure and unique properties upon exfoliation to a monolayer limit, make them particularly interesting. The band structure of 1T-$TiTe_2$ demonstrates the partial overlap of electron-like and hole-like bands leading to the metallic resistivity down to ~10K with the following increase of resistivity of a log-T type due to supposedly the two-dimensional Anderson localization in the weakly localized regime [1].

$TiSe_2$, the sister compound of $TiTe_2$, has gained significant attention due to its specific phase diagram as a function of pressure or intercalation by Cu. Both compounds assume the same crystal structure and exhibit the same physical properties at high temperatures. However, bulk $TiSe_2$ undergoes a CDW transition below 200 K, whereas $TiTe_2$ does not show a CDW state down to zero temperature. When Cu is intercalated or pressure is applied, the temperature of the CDW transition decreases and superconductivity appears [2]. Noteworthy, the CDW transition was observed in single-layer samples [3] and in epitaxial strained $TiTe_2$ multilayer films [4]. Apparently, the weak electron-phonon interaction (EPI) in $TiTe_2$ does not cause the appearance of superconductivity at ambient conditions. However, Dutta *et al.* [5] claim that small non-hydrostatic compression leads to an abrupt change in low-temperature resistance with a signature of possible CDW, followed by superconducting ordering at ∼5 GPa. Shortly, Zhou *et al.* [6] demonstrate that pressure-induced superconductivity can be correlated with the structural transition from the hexagonal P-3m1 phase to the monoclinic C2/m phase at ∼approximately 5.4 GPa, and the superconductivity remains robust up to 50.2 GPa.

Recently, we applied the Yanson point-contact (PC) spectroscopy method [7] to study the $TiSe_2$ and its doped modifications TiSeS and $Cu_xTiSe_2$ [8]. We observed the double maximum symmetrical relative to V=0 in *dV/dI* of $TiSe_2$ PCs, which narrows with increasing temperature and disappears above 170 K. These features correspond to the maximum on *ρ(T)* and are attributed to the CDW transition. Besides, we observed resistive switching (RS) in $TiSe_2$ as well as in TiSeS and $Cu_xTiSe_2$ PCs [8], between metallic-like low-resistance state (LRS) and semiconducting-like high-resistance state (HRS) with changing resistance up to two orders of magnitude at room temperature. A similar effect was observed also on other TMDs such as $MoTe_2$, $WTe_2$, $TaMeTe_4$ (Me = Ru, Rh, Ir) [9], $VS_2$ and $VSe_2$ [10], which could make this class of TMDs promising, e.g., for non-volatile ReRAM engineering, artificial neural networks, spintronics, and other nanotechnologies [11]. It was therefore challenging to apply Yanson's PC spectroscopy [7] also to the study of $TiTe_2$.

**Experimental details**

*Samples*. The single crystals of $TiTe_2$ were grown by a conventional chemical vapor transport method [12].

*Point-contact spectroscopy*. "Hard" PCs were prepared by touching a thin 100-150 μm Ag or Cu wire to the flat surface of a plate-like single-crystal flake that has been cleaved at room temperature or contacting its edge/side with this wire. Also, so-called "soft" PCs were made by dripping a small drop of silver paint between the sample edge and 50 μm thin Cu electrode. We measured the current-voltage *I(V)* characteristics of PCs and their first derivatives *dV/dI(V)*. The differential resistance *dV/dI(V)* was recorded by scanning the *dc* current *I* on which a small *ac* current *i* was superimposed using a standard lock-in technique. Thus, we conducted resistive measurements on the hetero-contacts between a normal metal (Ag, Cu) and investigated samples. The measurements were performed in the temperature range from liquid helium to room temperature. To observe the RS, we created PCs one by one at low temperatures inside the cryostat and then swept the voltage back and forth increasing its amplitude until the RS occurs.

**Results and discussion**

Figure 1 shows *dV/dI(V)* of PC on TiTe$_2$ with two symmetrically located with respect to zero bias maxima around +/- 150 mV, whose position narrows with increasing temperature and disappears above 150 K. A similar behavior was observed in *dV/dI* of sister compound TiSe$_2$ with CDW transition. In the latter case, these maxima in *dV/dI(V)* bear a resemblance to that observed in resistivity *ρ(T)*, which corresponds to the CDW transition. According to Dutta *et al.* [5], the semimetallic resistance *R* of TiTe$_2$ transforms to a humplike *R(T)* curve characteristic for CDW ordering at a non-hydrostatic pressure (about∼1.8 GPa). We assume that such pressure can be attained in "hard" PCs and that CDW ordering can be induced. We did not observe such characteristics in "soft" PCs, where pressure is unexpected. Additionally, non-reciprocal *dV/dI(V)* behavior is seen in Fig.1 by scanning voltage forth and back, likely as a precursor of RS observed in other PCs (see next Figures).

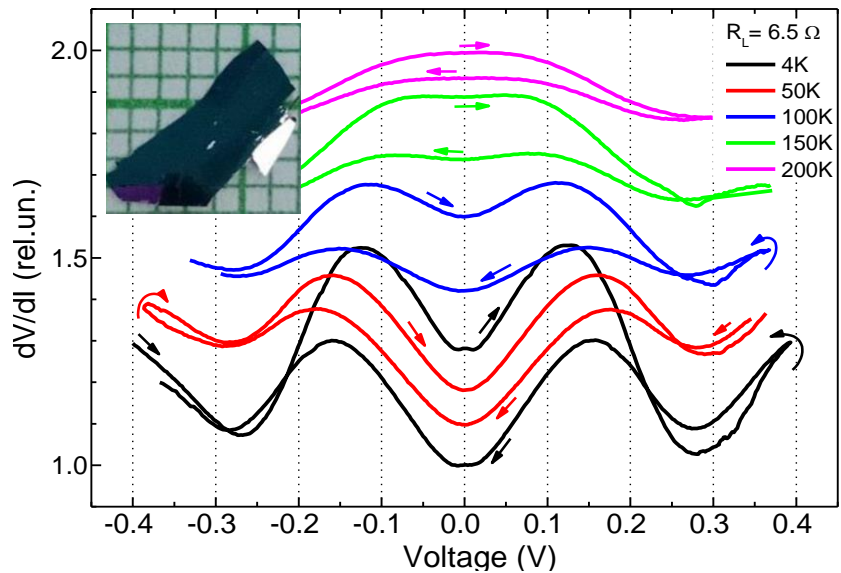

**Fig. 1.** *dV/dI(V)* recorded at the bias sweeping in both polarity directions of TiTe$_2$–Ag "hard" PC at different temperatures. Contact was established to the plane of TiTe$_2$. Arrows show the sweep directions of the current, back and forth. Note the hysteretic behavior of *dV/dI(V)*. All curves at temperatures above 4K are shifted upwards for clarity. The inset shows a TiTe$_2$ sample.

By analogy with TiSe$_2$, superconductivity would be expected in the vicinity of the CDW. However, we did not observe superconductivity in TiTe$_2$ PCs, probably because it needs either higher pressure or lower temperature.

It is known that the second derivative of *I(V)* characteristics of ballistic PCs contains straightforward information as to the EPI, i.e., the second derivative of *I(V)* or Yanson PC spectrum is directly proportional to EPI function α$^2$F(ω) [7]. As follows from theoretical calculations of phonon density of states F(ω) and Eliashberg EPI function α$^2$F(ω), the main phonon modes in TiTe$_2$ are located at 16, 22, and 33 meV [13]. Unfortunately, we have not yet been able to detect phonon features in our PC spectra of TiTe$_2$. As it can be seen from the left insert of Fig. 2, the second derivative $d(dV/dI)/dV(V)$ of TiTe$_2$ PC does not display any phonon features. According to Fig. 2 (left inset), only the position of the inflection in the second derivative roughly corresponds to the Debye energy of about 40 meV [13].

We should stress that mostly the backward scattering gives the largest contribution to the Yanson PC spectra of EPI [7]. However, as it was shown in Ref. [14], the backward scattering is substantially suppressed in layered systems with strong electronic correlations. For the mentioned reason, the observation of the phonon features in the Yanson PC spectra of layered TiTe$_2$ may be complicated. Another reason for the absence of phonon features on Yanson PC spectra is the realization of a thermal regime in PC [15, 16]. In this case, the phonon features in the spectra are considerably smeared out, which is accompanied by a steep background rise with voltage.

Figure 2 displays the observation of the RS in TiTe$_2$–Ag PC. When the voltage reaches about 200 mV at negative polarity, $dV/dI(V)$ jumps to about 600 mV, and then sweeping to positive polarity reveals a zero-bias peak instead of a broad minimum in the initial state. Then, at a positive voltage of around 350 mV, $dV/dI(V)$ returns to the LRS. Such a cycle with RS between LRS and HRS, with a change in resistance between them by one order of magnitude, is shown in Fig. 2. $I(V)$ curve demonstrates the "butterfly" shape at such switching (see middle inset in Fig. 2).

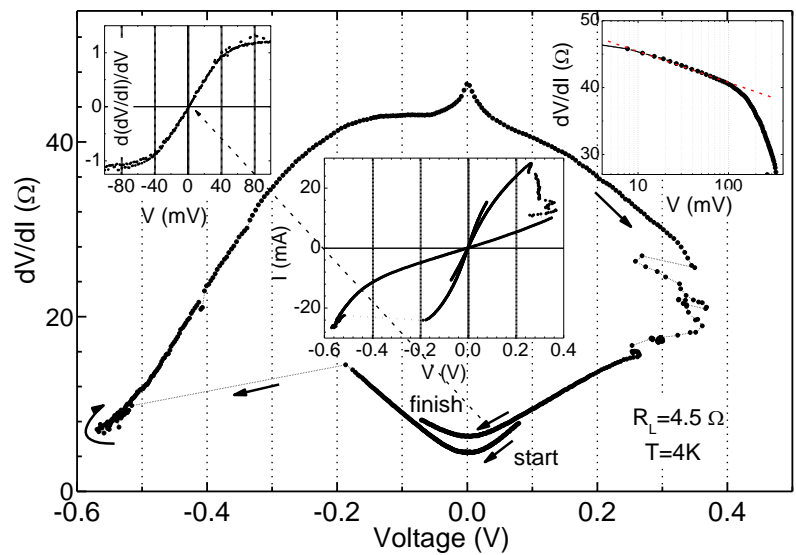

**Fig. 2.** Resistive switching in $dV/dI(V)$ for TiTe$_2$–Ag "hard" PC by "clockwise" sweeping of bias voltage. Inset (middle): $I(V)$ curve for the same PC. Inset (left): calculated/digital derivative of $dV/dI(V)$ from the main panel in LRS. Inset (right): $dV/dI(V)$ of HRS on log-scale. The red dashed line is a guide to the eye of logarithmic behavior.

Figure 3 displays RS for another PC at two temperatures. It can be seen that the RS effect, which is small-scale at 4K, increases significantly at 200 K (see $I(V)$ curves in the inset of Fig.3), probably due to the PC structure alteration. Usually, "hard" PCs are less stable by changing temperatures, and in particular, this PC did not survive when rising to room temperature to follow further development of the effect.

A question arises, what is the reason for nonmetallic $dV/dI(V)$ behavior in HRS? Let's remark that $dV/dI(V)$ obeys a log-$V$ dependence below 100 mV (see right inset in Fig. 2). It is reminiscent of log-$T$ dependence of $\rho(T)$ in TiTe$_2$ observed in Refs. [1, 6]. The authors attributed such behavior to the presence of disorder or Ti vacancies in the layers and self-intercalations of Ti between the layers. Therefore, the origin of the RS effect may be caused by a change of stoichiometry in the PC core due to the drift/displacement of Ti/Te vacancies under a high electric field.

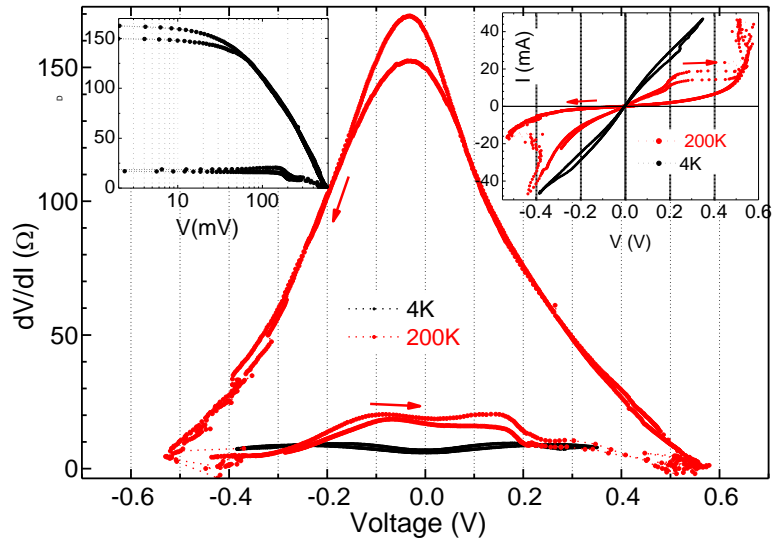

**Fig. 3.** Resistive switching in *dV/dI(V)* for TiTe$_2$–Ag "hard" PC at low and "high" temperatures. Inset (right): *I(V)* curves for PC from the main panel. Inset (left): *dV/dI(V)* at 200K in log-scale.

To better understand the observed RS effect, we must clarify a few experimental details. We prepared "hard" PCs as is written in the section Experimental details. The initial PC resistance is usually in the range of several to several tens of Ohms. Then we sweep voltage back and forth, increasing its amplitude from one cycle to the next one until a sudden resistance jump is observed at threshold voltage $V_{th}$. Then, we register the whole resistive loop (see e.g., Fig. 3) one or several times. That is, we started from LRS, contrary to typical cases in the literature, when RS is observed in the capacitor-like geometry, where a sample is placed between two electrodes (see, e.g. [11]). In the latter case, the formation of the LRS conductive channel is considered by applying to the sample in HRS a so-called "set" voltage up to several volts. This "set" or "forming" voltage creates a conductive channel/filament and is sufficiently larger than a threshold voltage $V_{th}$. On the contrary, in our case, PC is formed on one side of samples and we start from the metallic type LRS to look for the switching to HRS by approaching $V_{th}$. We assume that in our case reversible modification of the crystal structure in PC takes place under a high electric field. The resistance in HRS increases about an order of magnitude and demonstrates nonmetallic behavior with a sharp maximum at zero bias and log-*V* behavior at low temperatures (see Fig. 2).

Let us recall that we have observed a similar RS effect on a series of other TMDs [8, 9, 10] with van der Waals layered structure consisting of different transition metals such as Mo, W, Ta, Ti, Ru, Rh, Ir and chalcogenides such as S, Se, and Te. Furthermore, in every case, we observed *dV/dI* with log-*V* behavior in HRS at low temperatures. Therefore, we argue that the mechanism of RS must be similar for all these TMDs. The most common reason for log-*T* (log-*V* in the case of PCs) behavior in resistance at low temperatures is the well-known Kondo effect [7, 17] or weak localization in disordered 2-D systems [18, 19]. However, weak localization results only in a small correction to the resistance and is sensitive to a magnetic field, which doesn't seem to be the case for this study. Regarding the Kondo effect, it is unclear where the local magnetic moments come from when switching from LRS to HRS and why they disappear when switching back. Concerning Anderson localization, mentioned by Koike *et al.* in TiTe$_2$, it also adds only a few percent to the residual resistance at low temperatures [1]. Thus, the probable nature of the RS, as we mentioned above, is modification of the crystal structure in the PC core due to drift/displacement of Ti/Te ions/vacancies under a high electric field taking into account the strong variation of resistivity in $TiTe_2$ on stoichiometry. We have performed ab initio modeling of phase transition pathways to estimate how N, O, Se or Te atoms and different vacancies migrate inside $TiSe_2$ and $TiTe_2$. It turned out that the more probable candidates for interstitial atom migration are oxygen and tellurium. Our calculation shows that Ti interlayer ions have strong covalent bonding with nearest atoms and are well localized.

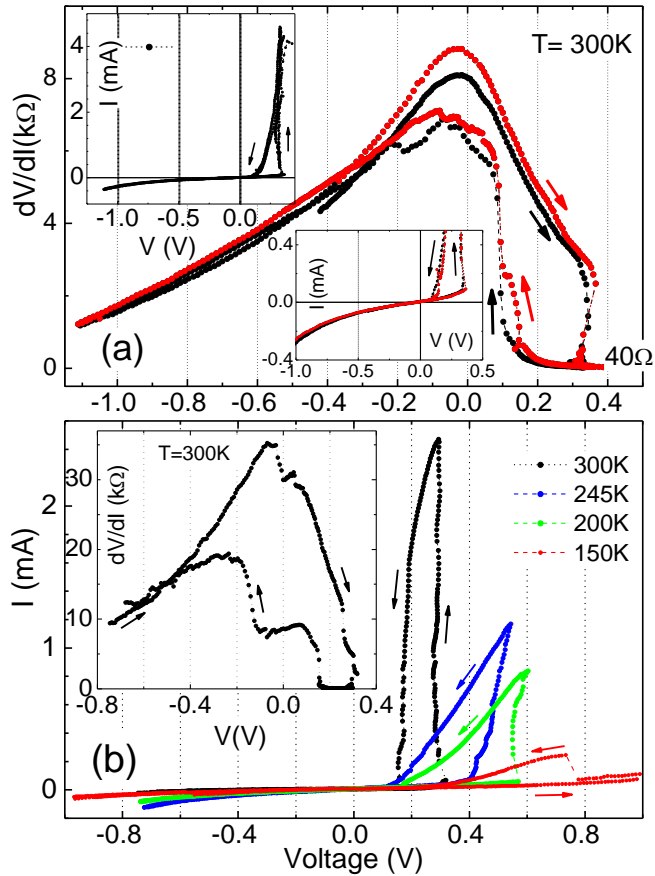

**Fig. 4.** (a) *dV/dI(V)* and *I(V)* curves (both insets) for $TiTe_2$ "soft" PC at 300K with diode-like effect. Black symbols correspond to the first sweeping cycle, red symbols to the next sweeping cycle. (b) Transformation of *I(V)* curves of another $TiTe_2$ "soft" PC with diode-like effect at lowering temperature. Inset: *dV/dI(V)* of *I(V)* curves at 300 K. Voltage polarity corresponds to silver paint.

Figure 4(a) shows *dV/dI(V)* and *I(V)* curves at 300K of "soft" $TiTe_2$ PC, where *I(V)* curve demonstrates diode-like behavior. The different symbol colours in Fig. 4(a) show the sweeping dynamics for the two cycles. It looks like the interface in a "soft" PC contains a semiconducting layer probably due to a degraded/oxidized/nonstoichiometric $TiTe_2$ surface and the formation of a Schottky-type barrier at the interface between this layer and the silver paint. Figure 4(b) demonstrates the transformation of *I(V)* curves of another "soft" $TiTe_2$ PC at a lower temperature in more detail. Note, here the resistance of PC at zero voltage in HRS (see inset) is at least twice as large as the maximal resistance of metallic (single atom) PC of about 13 kΩ [20], therefore, only the presence of a barrier can be the reason. Contrary to the sharp switching behavior at a definite voltage (see Figs. 2 &3), the hysteretic loop at a positive voltage (negative voltage at

TiTe$_2$ side) is formed in *I(V)* by gradual voltage sweep. It's more like a migration/diffusion of atoms/ions forming a barrier, which occurs under the action of an applied electric field/current. The vanishing of the diode effect with temperature decrease in Fig. 4(b) is in line with the mentioned migration/diffusion mechanism, which is slowed down by temperature decrease.

An interesting behavior of *dV/dI(V)* and *I(V)* was observed by decreasing temperature in Fig. 5. The shape of these curves varies from diode-like behavior at 140K to switching characteristics at low temperatures (see *dV/dI(V)* and *I(V)* at 2K). Besides, the switching amplitude in *dV/dI(V)* is high at 2K, contrary to PC in Fig. 3, where the RS effect vanishes at low temperatures. This is likely because the local arrangement of atoms in PC changes under temperature variation, local electric field change, thermal expansion/contraction, etc. It should be noted that during long-term measurements with changing temperatures, it is difficult to maintain stable PCs, therefore the data in Figs. 3, 4, and 5 are presented only for a few temperatures. Also, note log-*V* behavior of *dV/dI(V)* at 2K up to 300 mV (inset in Fig. 5a), as was mentioned above for other studied TMDs [8, 9, 10].

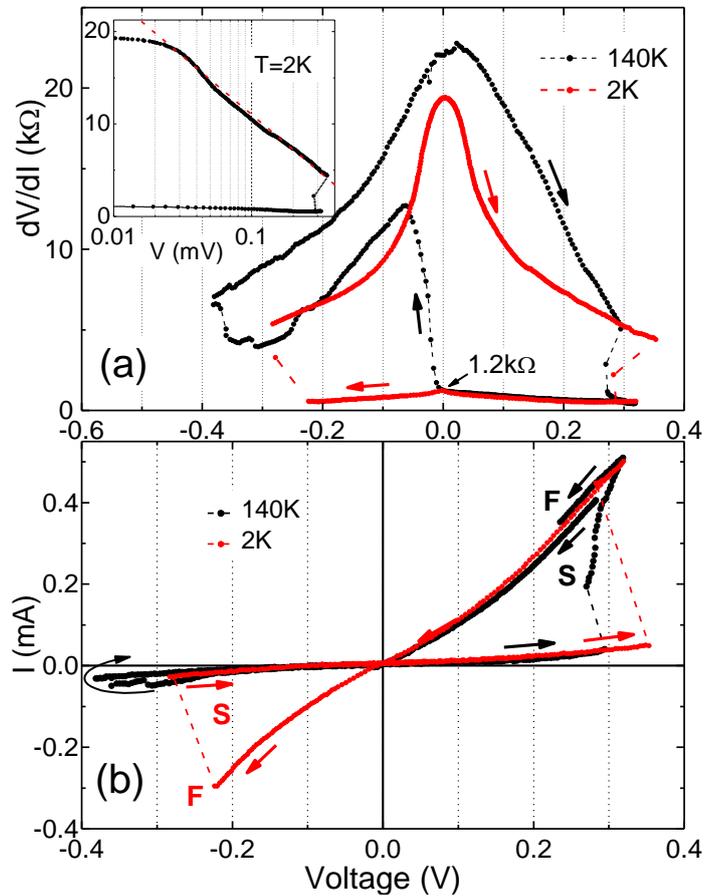

**Fig. 5.** (a) *dV/dI(V)* of TiTe$_2$ "soft" PC at two temperatures. Inset: *dV/dI(V)* in log-scale at 2K. The red dashed line is a guide to the eye of logarithmic behavior. (b) *I(V)* curves corresponding to the upper panel. S and F denote the start and finish of the voltage sweep, respectively.

Further, the diode effect is regularly observed in "soft" contacts, which likely consist from a number of contacts between microscopic Ag particles and sample. Contrary to the mechanical "hard" contacts, in this case, no mechanical stress or pressure is expected at PC creating, thus, if some nonmetallic thin surface layer (oxide) exists, then it prevents direct metallic contacts, and some Schottky–type barrier is created resulting in diode-like behavior. A curious loop in *I(V)* at positive voltage says that this barrier is an electric field/current dependent. Oddly, we did not observe a similar diode effect in "soft" PCs based on other TMDs [8, 9, 10].

**Conclusion**

We observed features in *dV/dI(V)* characteristics of TiTe$_2$ PCs, which indicate the emergence of CDW ordering below 150K likely due to the local pressure in the "hard" PC. They appear in the form of symmetrical relatively V=0 maxima in *dV/dI(V)*, similar to the case of typical CDW compound TiSe$_2$ [8]. At the same time, we did not observe superconducting features in our characteristics, probably, because it requires higher pressure or lower temperature. Besides, we detected RS in TiTe$_2$ PCs, between metallic-like LRS and semiconducting-like HRS with the change of resistance by an order of magnitude. Additionally, an unexpected diode-like effect was observed in "soft" TiTe$_2$ PCs with hysteretic *I(V)* at negative voltage on TiTe$_2$. Modification of the crystal structure in the PC core due to drift/displacement of Ti/Te ions/vacancies under a high electric field/current density is the more probable reason for the observed effects. In this way, discovering the RS and diode effect adds TiTe$_2$ to the list of layered TMD compounds, which can be promising, e.g., as building blocks for in-memory computing, to create innovative non-volatile ReRAM and other up-and-coming nanotechnologies. On the other hand, our observation and studies of the RS and diode effects in TMDs demonstrate the great potential of the Yanson PC spectroscopy method for the search for promising materials, which will help reveal the internal nature of these intriguing phenomena.

**Acknowledgment**

We appreciate V.V. Fisun, U. Nitzsche, and T. Schreiner for their technical assistance. We want to acknowledge Alexander von Humboldt's (OK, DE, BB, YuN) and Volkswagen Foundation's (OK) funding. OK and YuN are also grateful for support by the National Academy of Sciences of Ukraine under project Ф19-5 and are thankful to the IFW Dresden for its hospitality. L.H. acknowledges funding from the National Mission on Interdisciplinary CyberPhysical Systems (NM-ICPS) of the Department of Science and Technology, Government of India through the I-HUB Quantum Technology Foundation, Pune, India, for supporting research on 2D quantum material. OF gratefully acknowledges financial support from BMBF through the GU-QuMat project (01DK24008).